\documentclass[review]{elsarticle}
\usepackage{graphicx}
\usepackage{sidecap}

\journal{Journal of \LaTeX\ Templates}

\begin{document}

\begin{frontmatter}

\title{Time Projection Chamber (TPC) Detectors for \\ Nuclear Astrophysics Studies With Gamma Beams 
}


\author{M. Gai$^{1}$, D. Schweitzer$^1$, S.R. Stern$^1$, A.H. Young$^1$, 
R. Smith$^2$, M. Cwiok$^3$,  J.S. Bihalowicz$^3$, H. Czyrkowski$^3$, R. Dabrowski$^3$, W. Dominik$^3$, \\ A. Fijalkowska$^3$, Z. Janas$^3$, L. Janiak$^3$, A. Korgul$^3$, T. Matulewicz$^3$, \\ C. Mazzocchi$^3$, M. Pf\"{u}tzner$^3$, M. Zaremba$^3$, D. Balabanski$^4$, I. Gheorghe$^4$, \\ C. Matei$^4$, O. Tesileanu$^4$, N.V. Zamfir$^4$, M.W. Ahmed$^{5,6}$, S.S. Henshaw$^5$, \\ C.R. Howell$^5$, J.M. Mueller$^5$, L.S. Myers$^5$, S. Stave$^5$, C. Sun$^5$, H.R. Weller$^5$, Y.K. Wu$^5$,  A. Breskin$^7$, V. Dangendorf$^8$, K. Tittelmeier$^8$, M. Freer$^9$ }
\address{1. LNS at Avery Point, University of Connecticut, CT 06340, USA\\
 2.  Faculty of Arts, Computing, Engineering and Sciences, Sheffield Hallam University, Sheffield, S1 1WB, UK \\
 3. Faculty of Physics, University of Warsaw, Warsaw, Poland \\
 4.  Extreme Light Infrastructure-Nuclear Physics, Horia Hulubei National Institute for R\&D in Physics and Nuclear Engineering
Bucharest-Magurele, Romania\\
 5. Triangle Universities Nuclear Laboratory and Department of Physics, Duke University, Durham, NC 27708, USA \\
 6. Department of Mathematics and Physics, North Carolina Central University, Durham, NC 27707, USA\\
 7. Department of Particle Physics and Astrophysics, Weizmann Institute of Science, 76100 Rehovot, Israel \\
 8. Physikalisch-Technische Bundesanstalt, 38116 Braunschweig, Germany\\
 9. School of Physics and Astronomy, University of Birmingham, Edgbaston, Birmingham, B15 2TT, UK}




\begin{abstract}
Gamma-Beams at the HI$\gamma$S facility in the USA and anticipated at the ELI-NP facility, now constructed in Romania, present unique new opportunities to advance research in nuclear astrophysics; not the least of which is resolving open questions in oxygen formation during stellar helium burning via a precise measurement of the $^{12}$C($\alpha,\gamma$) reaction. Time projection chamber (TPC) detectors operating with low pressure gas (as an active target) are ideally suited for such studies. We review the progress of the current research program and plans for the future at the HI$\gamma$S facility with the optical readout TPC (O-TPC) and the development of an electronic readout  TPC for the ELI-NP facility (ELITPC). 
\end{abstract}

\begin{keyword}

Time Projection Chamber, Optical Readout, Electronic Readout, Gamma-Beams, Nuclear Astrophysics, Stellar Helium Burning
\end{keyword}

\end{frontmatter}


\section{Introduction}

\paragraph{Gamma-Beams} Gamma-beams (2--20 MeV) proved to be enormously useful for low energy nuclear physics studies in the pioneering work at the High Intensity Gamma-ray Source (HI$\gamma$S) facility at the Triangle Nuclear Physics Laboratories (TUNL) located at Duke University in the USA \cite{HIgS}. Further improvement of the energy resolution (by a factor 5) and intensity (by a factor of 10)  anticipated for the Extreme Light Infrastructure -- Nuclear Physics (ELI-NP) facility under construction at Magurele near Bucharest in Romania \cite{ELI-NP}, promises to allow some of the most crucial measurements in nuclear astrophysics. Specifically, the C/O ratio at the end of stellar helium burning has been emphasized as a problem of {\em ``paramount importance"} in nuclear astrophysics \cite{Fow84}. To solve this problem we need to measure with high accuracy the cross section of the $^{12}$C($\alpha,\gamma$) reaction at low energies approaching center of mass energy of 1.0 MeV and resolve the nagging ambiguities \cite{Gai13} in the extrapolated values of the p-wave and d-wave cross sections at the Gamow window (300 keV), designated by S$_{E1}$(300) and S$_{E2}$(300), correspondingly. The high intensity and improved energy-resolution anticipated for the gamma-beam of the ELI-NP provides a unique opportunity for a high precision measurement of the $^{12}$C($\alpha,\gamma$) reaction at E$_{cm}$  = 1.1 MeV by measuring the inverse $^{16}$O($\gamma,\alpha$) reaction with a gamma-beam of E$_\gamma$ = 8.26 MeV. A detailed and complete angular distribution spanning the entire angular range of 0--180$^\circ$ appears possible with a three-week measurement \cite{RomReport}. Such a complete angular distribution spanning the entire angular range of 0--180$^\circ$ was demonstrated \cite{WRZ14} to permit a separation of the E1 and E2 cross sections and the corresponding mixing phase-angle ($\phi_{12}$) with very high precision.
    
\paragraph{Optical TPC Detectors} Optical readout tme projection chamber (TPC) detectors were used in pioneering measurements with radioactive beams at the NSCL in MSU \cite{Cwiok0,Cwiok} and with gamma-beams at the HI$\gamma$S facility \cite{WRZ14,Gai10}. But these detectors use optical readout and suffer from low counting rates that does not permit the use of the full beam power (even at the HI$\gamma$S setup). In order to fully utilize the high intensity anticipated for the ELI-NP gamma beam an electronic readout (eTPC) concept has been developed \cite{RomReport}. A mini-TPC \cite{Cwiok18} prototype of the full detector planned to be used at the ELI-NP facility (ELITPC) was constructed at the University of Warsaw and delivered to the ELI-NP facility. 

\section{Measurements at the HI$\gamma$S With the O-TPC}

\paragraph{Data on the $^{16}$O($\gamma,\alpha$) reaction} A large volume of data (approximately 4 TB) collected at the HI$\gamma$S facility are now being analyzed \cite{Robin,Sarah}. The goal of this analysis is two fold: First, we plan to extract with high precision the E1 and E2 cross sections and the E1-E2 mixing phase angle ($\phi_{12}$) measured in the angular distribution of the $^{16}$O($\gamma,\alpha$) reaction with gamma-beams at E$_\gamma$ = 9.08, 9.39, 9.58, 9.78 MeV with two different gas mixtures CO$_2$(80\%) + N$_2$(20\%) and N$_2$O (80\%) + N$_2$(20\%), at 100 Torr. Second, the development of the analyses routine will prepare us to analyze the data anticipated from measurements with the ELITPC. After completing the current HI$\gamma$S data analyses we plan to use the 200 hours of already approved beam time to measure at the lowest possible energy at the HI$\gamma$S facility (constrained by the count-rate) at E$_\gamma$ = 8.80 MeV.

\begin{figure*}
\includegraphics[width=4.5in,clip]{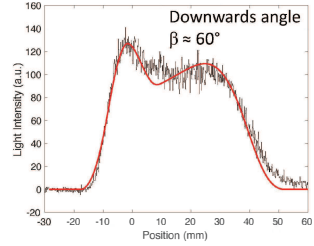}
\caption{(color online) An $^{16}$O dissociation event: the PMT time projection signal (on right) and the CCD image with its longitudinal projection (on left) \cite{Robin}. The PMT time signal is converted to distance from the interaction point using the measured drift velocity of 11.95 mm/$\mu$s and the fit of the line shape using dE/dx is shown in red line.}
\label{Robin}       
\end{figure*}

\paragraph{Measurement with N$_2$O gas mixture} The use of N$_2$O gas mixture which was developed at the Weizmann Institute \cite{Leo} proved to be very beneficial since it removes the background from the $^{12}$C($\gamma,3\alpha$) reaction. But the poor energy resolution obtained with the N$_2$O gas mixture (most likely due to e-N$_2$O resonance) prevented the use of the anode grid signal to measure the total energy deposited and necessitated relying on the measured track length to separate reactions from the dissociation of $^{16}$O and $^{18}$O with $\Delta$Q = 935 keV. In Fig.~\ref{Robin} we show a typical event measured with the N$_2$O gas mixture including the track recorded by the CCD camera \cite{Gai10} and the time projection PMT signal \cite{Gai10}. We are currently analyzing all available data to extract the measured angular distributions with the N$_2$O gas mixture \cite{Robin}.

\section{The ELITPC Detector}
\paragraph{The proposed ELITPC detector} The ELITPC detector proposed by the charged-particle working group \cite{RomReport} has been reviewed by the ELI International Scientific Advisory Board (ISAB) and was approved for construction and installation at the ELI-NP. Briefly, it utilizes an electronic readout in the horizontal plane perpendicular to the drifting electrons, of three lines oriented at 120$^\circ$ to each other placed on a multi-layer PC board (commonly referred to as a u-v-w readout). The electron multiplication is achieved with three 35x20 cm$^2$ Gas Eelectron Multipliers (GEMs). The ELITPC is proposed as one of the two main detectors for measurement of charged particles of relevance to nuclear astrophysics as discussed in \cite{RomReport}.

\begin{SCfigure}
\includegraphics[width=2in,clip]{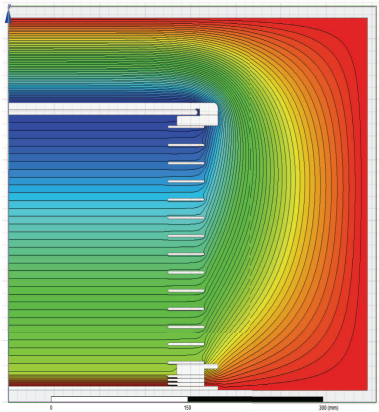}
\caption{(color online) MAXWELL simulation of the ELITPC detector \cite{Deran} showing the 15 mm spaced 40 mm wide field guard electrodes that result in equally spaced (by 285 V) parallel equipotential lines. These equipotential lines indicate a uniform electric field in the drift cage between the shown cathode and anode.}
\label{Deran}       
\end{SCfigure}

\paragraph{The mini-TPC Prototype Detector} A smaller mini-TPC detector has been constructed at the University of Warsaw \cite{Cwiok18} in order to study and optimize the performance characteristics of the ELITPC. The homogeneity of the electric field was simulated using MAXWELL \cite{Deran} to be better than 1 V/cm ($<$0.5\%) and the results are shown in Fig.~\ref{Deran}.

\paragraph{Test of the mini-TPC at the IFIN Tandem} The mini-TPC was tested with alpha-beams extracted from the IFIN tandem as well as with neutrons produced by the same alpha-beam with a Be target. In Fig.~\ref{IFIN} we show an event of $^{16}$O dissociation by a neutron vividly displaying the reconstructed alpha-particle and $^{12}$C tracks.

\section{Conclusion}
\begin{SCfigure}
\includegraphics[width=2.5in,clip]{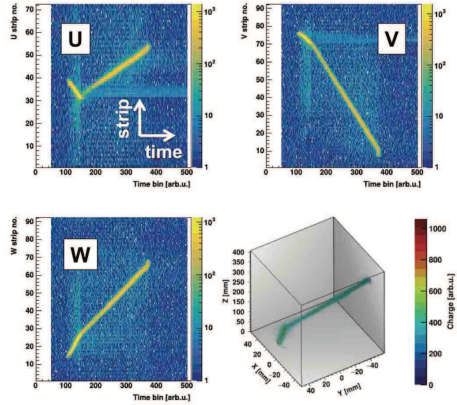}
\caption{(color online) Raw signals collected on U, V and W coordinates and 3D reconstruction of the reaction products of neutron interaction with the CO$_2$ gas at 100 mbar \cite{Cwiok18a}.}
\label{IFIN}       
\end{SCfigure}

In the analyses of the HI$\gamma$S data, $^{16}$O and $^{18}$O dissociation events have been identified and differentiated from background and angular distributions are being generated. The first measurement of $^{16}$O dissociation at the IFIN shown in Fig.~\ref{IFIN} serves as a ``proof of principle" of the design specs of the ELITPC which is now ready to move to the construction and installation phase at the ELI-NP. 

\section{Acknowledgements}
The material presented in this paper is based upon work supported by the U.S. Department of Energy, Office of Science, Nuclear Physics, Award No. DE-FG02-94ER40870 and DE-FG02-91ER-40608. Scientific work is supported by the Polish Ministry of Science and Higher Education from the funds for years 2017--2018 dedicated to implementing the international co-funded project no. 3687/ELI-NP/2017/0 and by ELI-NP/IFIN-HH under the Collaborative R\&D Project Agreement no. 88/25.10.2016. The ELI-NP team acknowledges the support from the Extreme Light
Infrastructure Nuclear Physics (ELI-NP) Phase II, a project co-financed by the Romanian Government and the European Union through the European Regional Development Fund - the Competitiveness Operational Programme (1/07.07.2016, COP, ID 1334).





\section{References}

\begin{thebibliography}{9}

\bibitem{HIgS} Henry R. Weller, Mohammad W. Ahmed, Haiyan Gao, Werner Tornow, Ying Wu, Moshe Gai, Rory Miskimen, Prog. Part. Nucl. Phys. {\bf 62}, 257 (2009).


\bibitem{ELI-NP} D. Filipescu, A. Anzalone, D.L. Balabanski, S. S. Belyshev, F. Camera, M. La Cognata, P. Constantin, L. Csige, P. V. Cuong, M. Cwiok, V. Derya, W. Dominik, M. Gai, S. Gales, I. Gheorghe, B.S. Ishkhanov, A. Krasznahorkay, A. A. Kuznetsov, C. Mazzocchi, V.N. Orlin, N. Pietralla, M. Sin, K.A. Stopani, O. Tesileanu, C.A. Ur, I. Ursu, H. Utsunomiya, V.V. Varlamov, H.R. Weller, N.V. Zamfir, and A. Zilges, Eur. Phys. J. A {\bf 51}, 185 (2015).
 

\bibitem{Fow84} W.A. Fowler, Rev. Mod. Phys. {\bf 56}, 149 (1984).

\bibitem{Gai13} Moshe Gai, Phys. Rev. C {\bf 88}, 062801(R) (2013).


\bibitem{RomReport} O. Tesileanu, M. Gai, A. Anzalone, C. Balan, J.S. Bihalowicz, M. Cwiok, W. Dominik, S. Gales, D.G. Ghita, Z. Janas, D.P. Kendellen, M. La Cognata, C. Matei, K. Mikszuta, C. Petcu, M. Pf\u..tzner, T. Matulewicz, C. Mazzocchi, C. Spitaleri, Rom. Rep. Phys. {\bf 68}, S699 (2016).

\bibitem{WRZ14} W.R. Zimmerman, M.W. Ahmed, B. Bromberger, S.C. Stave, A. Breskin, V.Dangendorf, Th. Delbar, M. Gai, S.S. Henshaw, J.M. Mueller, C. Sun, K. Tittelmeier, H.R. Weller and Y.K. Wu, Phys. Rev. Lett. {\bf 110}, 152502 (2013).

\bibitem{Cwiok0} M. Cwiok, W. Dominik, Z. Janas, A. Korgul, K. Miernik, M. Pf\"{u}tzner, M. Sawicka, and A. Wasilewski;  IEEE Trans. Nucl. Sci., {\bf 52,\#6}, 2895 (2005).

\bibitem{Cwiok} K. Miernik, W. Dominik, Z. Janas, M. Pf\"{u}tzner, L. Grigorenko, C. R. Bingham, H. Czyrkowski, M. Cwiok, I. G. Darby, R. Dabrowski, T. Ginter, R. Grzywacz, M. Karny, A. Korgul, W. Kusmierz, S. N. Liddick, M. Rajabali, K. Rykaczewski, and A. Stolz, Phys. Rev. Lett. {\bf 99}, 192501 (2007).

\bibitem{Gai10} M. Gai, M.W. Ahmed, S.C. Stave, W.R. Zimmerman, A. Breskin, B. Bromberger, R. Chechik, V. Dangendorf, Th. Delbar, R.H. France III, S.S. Henshaw, T.J.  Kading, P.P. Martel, J.E.R. McDonald, P.-N. Seo, K. Tittelmeier, H.R. Weller,  and A.H. Young, JINST {\bf 5}, 12004 (2010).


\bibitem{Cwiok18} M. Cwiok, M. Bieda, J.S. Bihalowicz, W. Dominik, Z. Janas, L. Janiak, J. Manczak, T. Matulewicz, C. Mazzocchi, M. Pf\"{u}tzner, P. Podlaski, S. Sharma, M. Zaremba, D. Balabanski, A. Bey, D.G. Ghita, O. Tesileanu, M. Gai, Acta Phys. Pol. B {\bf 49}, 1001 (2018).


\bibitem{Robin} Robin Smith, M. Gai, M.W. Ahmed, M. Freer, I. Gheorghe, C.R. Howell, S.R. Stern, IOP Nucl. Phys. Conf. 4th April 2018, University of the West of Scotland.

\bibitem{Sarah} Sarah R. Stern, Moshe Gai, Robin Smith, Mohammad W. Ahmed, Calvin R. Howell, \\ National Nucl. Phys. Summer School, Yale, New Haven, June 18 - 29, 2018.

\bibitem{Leo} L. Weissman, M. Gai, A. Breskin, R. Chechik, V. Dangendorf, K. Tittelmeier, H.R. Weller, JINST {\bf 1}, 05002 (2006).


\bibitem{Deran} Deran Schweitzer, Moshe Gai, Mikolai Cwiok, Wojciech Dominik, Mikhael Guy, Dimiter Balabanski, Catalin Matei, \\ National Nucl. Phys. Summer School, Yale, New Haven, June 18 - 29, 2018.

\bibitem{Cwiok18a} M. Cwiok, J.S. Bihalowicz, H. Czyrkowski, R. Dabrowski, W. Dominik, A. Fijalkowska, Z. Janas, L. Janiak, K. Kierzkowski, A. Korgul, T. Matulewicz, C. Mazzocchi, W. Oklinski, M. Pf\"{u}tzner, M. Zaremba, D. Balabanski, A. Bey, D.G. Ghita, L. Guardo, C. Matei, M. Gai, D. Schweitzer, \\ Nuclear Photonics 2018, June 24 -- 29, 2018, Brasov, Romania.

\end{thebibliography}

\end{document}